\documentclass[aps,prb,groupedaddress,twocolumn,showpacs]{revtex4-1}
\usepackage{amsmath}
\usepackage{amsthm, amssymb} 
\usepackage{bbold}                       
\usepackage{graphicx}
\usepackage{color}
\usepackage{times, verbatim}      

\usepackage{natbib}
\usepackage{hyperref}
\hypersetup{
       colorlinks=true,
}

\begin{document}

\title{The enigmatic 12/5 fractional quantum Hall effect} 

\author{Kiryl Pakrouski$^{1}$, Matthias Troyer$^{1,2,3}$, Yang-Le Wu$^{4}$, Sankar Das Sarma$^{4}$  and Michael R. Peterson$^{5}$}
\affiliation{$^{1}$Theoretische Physik and Station Q Zurich, ETH Zurich, 8093 Zurich, Switzerland}
\affiliation{$^{2}$Quantum Architectures and Computation Group, Microsoft Research, Redmond, WA, USA}
\affiliation{$^{3}$Microsoft Research Station Q, Santa Barbara, CA, USA}
\affiliation{$^{4}$ Joint Quantum Institute and Condensed Matter Theory Center, Department of Physics, University of Maryland, College Park, Maryland 20742, USA}
\affiliation{$^{5}$Department of Physics \& Astronomy, California State University Long Beach,  Long Beach, California 90840, USA}

\begin{abstract}
We  numerically study the fractional quantum Hall effect at filling factors $\nu=12/5$ and 13/5 (the particle-hole conjugate of 12/5) in high-quality two-dimensional GaAs heterostructures via exact diagonalization including finite well width and Landau level mixing. We find  that Landau-level mixing suppresses the  $\nu=13/5$ fractional quantum Hall effect relative to  $\nu=12/5$.  By contrast, we find  both $\nu=2/5$ and (its particle-hole conjugate) $\nu=3/5$ fractional quantum Hall effects in the lowest Landau level to be robust under Landau-level mixing and finite well-width corrections. Our results provide a possible explanation for the experimental absence of the 13/5 fractional quantum Hall state as caused by Landau-level mixing effects.
\end{abstract}

\date{\today}


\maketitle

\section{Introduction}
There is  interest across physics, mathematics, engineering, materials research, and computer science in finding robust experimental manifestations of topologically ordered phases with non-Abelian anyonic low-energy excitations. Not only are non-Abelian anyons (i.e., neither fermions nor bosons) suitable for topological quantum computation, but they are  described by topological quantum field theories (TQFTs) of intrinsic fundamental interest~\cite{Nayak08}. The fractional quantum Hall effect~\cite{Tsui82,DasSarma97,jain2007composite} (FQHE) is the canonical example of a system supporting topologically ordered phases and is widely thought to support non-Abelian anyons in the second orbital electronic Landau level (LL), most probably at filling factor $\nu=5/2$~\cite{Willett87}.  There is a  possibility that  the experimentally observed FQHE at $\nu=12/5$ supports particularly exotic topologically ordered phases described by the $Z_3$ parafermionic Read-Rezayi  states~\cite{Read99,RR2009,Bonderson2012,ZhuHaldane2015,ZaletelMongArxiv2015,Geraedts2015,Peterson2015,Zhao2015},  exemplifying an exotic SU$(2)_3$ TQFT  [in contrast to the 5/2 FQH state belonging to the SU$(2)_2$ TQFT].   Since SU$(2)_3$ TQFT supports a richer version of non-Abelian anyons that can realize \textit{universal} fault-tolerant quantum computation~\cite{Nayak08}, there is a great deal of interest in the 12/5 FQHE.  In this work, we focus on the enigmatic FQHE at $\nu=12/5$.

Compared to the rather ubiquitous $\nu=5/2$ FQHE, the experimental literature for $\nu=12/5$ (= 2 + 2/5 filling) is  sparse with only a few experimental reports of its observation.  The 12/5 FQHE  was observed in a 30 nm wide GaAs quantum well with  electron densities of $n\sim 3\times 10^{11}$cm$^{-2}$ at  magnetic field strengths of  $B\sim 5$ Tesla at temperatures $T\sim 6$-$36$ mK~\cite{Xia04,Kumar10,Choi08,Pan08,Zhang2012,Gabor12}.  In addition to its fragility (the 12/5 FQHE is observed only in the highest quality samples with little disorder), the real enigma  is  the corresponding particle-hole conjugate FQHE at 13/5 ($=5-12/5$), which has never been observed in spite of other FQHEs in the second LL (e.g., 7/3 and 8/3, 11/5 and 14/5) showing both particle-hole conjugate states with roughly equal strength.  This discrepancy is  puzzling because in the lowest LL the FQHEs at $\nu=2/5$ and 3/5 are both routinely observed, are to  good approximation particle-hole conjugates of one another~\cite{Du93,Manoharan1994,PanPriv16}, and are well-described by the composite fermion (CF) theory~\cite{Jain89,jain2007composite}. The exotic, rather than CF-like nature of the 12/5 state has been discussed based on the analysis of the experimentally measured energy gap~\cite{Kumar10}. Interestingly, the 12/5 and 13/5 FQHEs (with roughly equal strength) are observed in systems where two subbands are occupied (e.g., bilayers, thick quantum wells) such that the chemical potential is  in the lowest LL (but in the higher subband  so  two LLs are completely full)~\cite{Liu11,YangLiu2015,Shabani2010}. In this work 
 we provide a possible explanation for the absence (presence) of a 13/5 (12/5) FQHE in the second LL as arising from the LL mixing effect that explicitly breaks the particle-hole symmetry.

Several candidate wave functions for $\nu=12/5$ have been proposed and studied~\cite{Bonderson2012,ZhuHaldane2015,ZaletelMongArxiv2015} under idealized conditions, using the  Coulomb interaction without particle-hole symmetry breaking.  Two recent numerical studies~\cite{ZhuHaldane2015,ZaletelMongArxiv2015} reinforced initial results~\cite{Read99,RR2009} that the ground state at $\nu=12/5$ is in the non-Abelian $Z_3$ Read-Rezayi (RR) phase. Both studies perturbed the interaction finding a finite region of stability around the Coulomb point.  All works considered particle-hole symmetric two-body Hamiltonians, so all conclusions made therein regarding the $\nu=12/5$ state are equally valid for the particle-hole conjugate state at $\nu=13/5$.  
Thus, existing theories provide evidence that the experimentally observed 12/5 and (unobserved) 13/5 FQHEs are both in the RR $Z_3$ phase, but cannot explain \textit{why} one (i.e., 12/5) exists experimentally and the other (i.e., 13/5) does not.
We provide a plausible explanation for this puzzle.

LL mixing breaks particle-hole symmetry through emergent three-body (and higher) terms  in an effective realistic Hamiltonian~\cite{Bishara09a,Peterson13b,Sodemann13,Rezayi13}.  The importance of LL mixing can be parameterized by the ratio $\kappa$ of the Coulomb energy $e^2/\epsilon l_0$ to the bare cyclotron energy $\hbar\omega$ (i.e., the LL gap): $\kappa=(e^2/\epsilon l_0)/\hbar \omega$, where $\epsilon$ is the background lattice dielectric constant, $l_0=\sqrt{\hbar c /eB}$ is the magnetic length, $e$ is the electron charge, and $\omega = eB/mc$ is the cyclotron frequency.  For GaAs, $\kappa\approx 2.5/\sqrt{B[\mathrm{T}]}$.  For most experiments in the second LL, $\kappa$ is of the order of unity, making LL mixing an important correction. One attempt at incorporating LL mixing  at $\nu=12/5$ used the  approximation of including  additional basis states within exact diagonalization~\cite{Wojs2009}, but did not investigate 13/5.

In the present work, we numerically study a realistic model of the FQHE in the second LL using exact diagonalization, systematically including LL mixing effects due to (the infinite number of) all other LLs.  We find that the LL mixing-induced particle-hole symmetry breaking strongly favors the $\nu=12/5$ FQHE over the 13/5 in the second LL, qualitatively in agreement with experimental observations.  By contrast, in the lowest LL, we do not find significant particle-hole symmetry breaking between  $\nu=2/5$ and 3/5 FQHE.  
Our work  gives a probable explanation for 
the presence (absence) of 12/5 (13/5) in the second LL 
and the existence and equal strength of 2/5 and 3/5 FQHEs in the lowest LL.  
Our work also strengthens the claim that at finite LL mixing, a 12/5 FQHE arises from a RR parafermionic non-Abelian state (rather than from Abelian composite fermion states as for the 2/5 and 3/5 FQHEs).

\section{Effective Hamiltonian}
Our realistic effective Hamiltonian describes $N_e$ interacting electrons confined to the $N^\mathrm{th}$ LL of a quasi-two-dimensional quantum well (modeled as an infinitely deep square well of width $w$) and incorporates LL and subband mixing.  The Coulomb interaction  causes virtual electron/hole excitations to higher/lower LLs and subbands included perturbatively to lowest order in $\kappa$ (note  this involves coupling all LLs~\cite{Peterson13b}). The effective Hamiltonian is
\begin{eqnarray}
H(w/\ell_0,\kappa,N)&=&\sum_{m}V^{(N)}_{\mathrm{2body},m}(w/{\ell_0},\kappa)\sum_{i<j}\hat{P}_{ij}(m)\nonumber\\
 &&\hspace{-1cm}+ \sum_{m} V^{(N)}_{\mathrm{3body},m}(w/{\ell_0},\kappa)\sum_{i<j<k}\hat{P}_{ijk}(m)
\label{Heff}
\end{eqnarray}
where $\hat{P}_{ij}(m)$ and $\hat{P}_{ijk}(m)$ are two- and three-body projection operators onto pairs  or triplets of electrons with
relative angular momentum $m$.  $V^{(N)}_{\mathrm{2body},m}(w/{\ell_0},\kappa)$ 
and $V^{(N)}_{\mathrm{3body},m}(w/{\ell_0},\kappa)$ are the two- and three-body effective pseudopotentials~\cite{Haldane83,Simon07c} in the $N^\mathrm{th}$ LL.
The full calculation of the two- and three-body pseudopotentials is quite involved and is given in detail in Ref.~\onlinecite{Peterson13b} for systems with finite thickness,  in Ref.~\onlinecite{Sodemann13} for zero thickness, and in Ref.~\onlinecite{Rezayi13} where the calculation is done numerically.  Here we provide a brief outline of the main details and encourage the reader to consult the above references.

In the absence of Landau level (LL) mixing, the planar pseudopotentials $V^{(N)}_{\mathrm{2body},m}$ can be calculated as  (see, for instance, Ref.~\onlinecite{jain2007composite})
\begin{equation}
\label{eq:VmfromVq}
V^{(N)}_{\mathrm{2body},m}
= \int_0^{\infty} q dq V(q) [L_N(q^2/2)]^2L_m(q^2)e^{-q^2},
\end{equation}
where $N$ is the LL index, $L_N$ are the Laguerre polynomials, and 
\begin{equation}
V(q) = \frac{1}{2\pi}\int d^2\mathbf{r}e^{i\mathbf{q}\cdot\mathbf{r}}V(r)
\end{equation}
is the Fourier transform of the real-space interaction potential $V(r)$.   For the case of finite width, $V(q)$ can be written as
\begin{equation}
\label{eq:VqEff}
V(q) =  \frac{e^2}{\epsilon q}\int dz_1\int dz_2 |\eta(z_1)|^2|\eta(z_2)|^2 e^{-q|z_1-z_2|},
\end{equation}
where $\eta(z)$ is the electron wave function in the $z$-direction.  For a realistic experimental system, $\eta(z)$ can be determined from solving the Schrodinger and Poisson equations self-consistently (see Ref.~\onlinecite{jain2007composite} for more details).  In this work, we consider an infinitely deep square well of width $w$ to model finite thickness, hence, $\eta(z) = \sqrt{2/w}\sin(\pi z/w)$.  

Pseudopotentials describing the pure Coulomb interaction can be derived in both the spherical and planar geometries.  Because the planar pseudopotentials do not depend on the system size it is more convenient to compute the pseudopotentials that include effects of finite thickness and Landau-level mixing in the planar geometry.  The spherical pseudopotentials extrapolate to the planar pseudopotentials in the limit of a sphere of infinite radius, i.e., the thermodynamic limit.  Further, it has been demonstrated that using planar pseudopotentials in the spherical geometry does not lead to qualitative differences compared with using spherical 
pseudopotentials (see for example Ref.~\onlinecite{Pakrouski2015}).

Beyond renormalizing the two-body interactions, LL mixing produces particle-hole symmetry breaking three-body terms (cf. Ref.~\onlinecite{Peterson13b}).  Equation~(\ref{Heff}) has a well-defined  exact limit as $\kappa\rightarrow 0$, hence, we  can  determine the leading-order effects of LL mixing on the FQHE.  Most experimental observations  of the 12/5 FQHE  occur at fields of $B\sim5.15$ T (see Ref.~\onlinecite{Kumar10}), giving a quantum well width (30 nm) of $w/l_0\approx 2.65$ and $\kappa\approx 1.1$.  We estimate (an exact self-consistent calculation is possible for a particular device~\cite{Pakrouski2015}) that an infinitely deep quantum well of $w/l_0\approx3$ provides approximately the same confinement as the real quantum well, and we consider $w/l_0\leq 4$ and $\kappa\neq 0$ to model realistic samples under realistic conditions.  
We assume fully spin-polarized~\cite{ZaletelMongArxiv2015} single-component states throughout this work. We consider $V^{(1)}_{\mathrm{3body},m}$ for $3\leq m\leq 8$-- previous work  demonstrated  that $m>9$ terms are unlikely to produce qualitative effects~\cite{Pakrouski2015}, especially for small $\kappa$.

We use the spherical geometry~\cite{Haldane83,jain2007composite} with the total magnetic flux  $N_\phi = N_e/f - S$, and where $f$ is the filling factor, as $N_e\rightarrow\infty$, of the $N^\mathrm{th}$  LL and $S$ is the shift~\cite{Wen90b}.  The experimental filling factor is $\nu=f + 2N$, where  2$N$ arises from completely filling the lower $N$ spin-up and -down LLs.  FQHE states are gapped uniform density ground states with  total angular momentum $L=0$.   The RR $Z_3$ state describes $f=3/5$ with $S=3$, while the particle-hole conjugate RR state,  conj($Z_3$), describes $f=2/5$ with $S=-2$.  The CF states for $\nu=2/5$ and 3/5  have shifts of $S=4$ and $-1$, respectively. 
Although the pairs of particle-hole conjugate states appear at different shifts, in the absence of LL mixing ($\kappa=0$) they have identical spectra and all eigenstates are particle-hole conjugates of each other. Hence, by considering properties such as energy gaps, overlaps, and entanglement spectra, we can isolate the effects of LL mixing.

\begin{figure}[]
 \includegraphics[width=8.cm,angle=0]{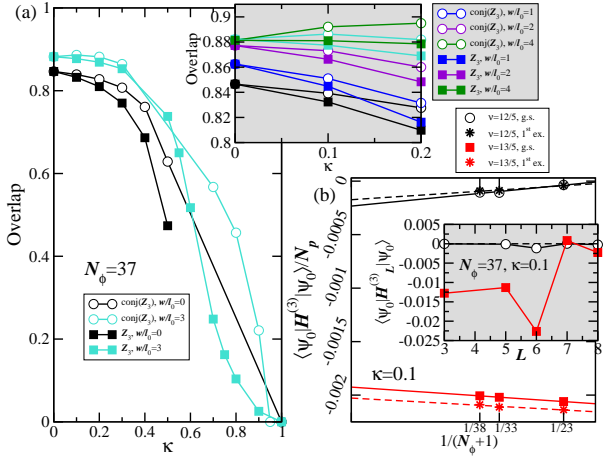}
  \caption{(Color online)  (a) Wave-function overlap between  $Z_3$  and conj($Z_3$)  and the exact ground state of Eq.~(\ref{Heff}) at $\nu=13/5$ and 12/5, respectively, as a function of $\kappa$ for $N_{\phi}=37$ (14 holes/electrons). A finite well width increases the overlaps and $\kappa$ breaks particle-hole symmetry, yielding  higher overlaps with  conj($Z_3$)  for 12/5 compared to  $Z_3$ for 13/5.  The inset shows the overlaps in more detail.
    (b)
    Expectation values of the three-body terms per particle $N_p$ of Eq.~(\ref{Heff}) for $\kappa=0.1$ and $w/l_0=0$, evaluated for the ideal Coulomb ground and first excited states (both denoted $|\psi_0\rangle$) at 12/5  and 13/5, respectively, as a function of inverse LL degeneracy [$1/(N_\phi+1)$] extrapolated to the thermodynamic limit. $N_\phi=27$ is aliased with $\nu=1/3$ and left out. Inset: Expectation values for each three-body term  [$H^{(3)}_L=V^{(N)}_{\mathrm{3body},L}(w/{\ell_0},\kappa=0.1)\sum_{i<j<k}\hat{P}_{ijk}(L)$] for $N_\phi=37$.  Lines are a guide to the eye, except in the main plot of (b) where they represent linear extrapolations. }
       \label{fig:z3overlaps_H3expValPlanarCoul}
\end{figure}

\begin{figure}[]
  \includegraphics[width=9.cm,angle=0]{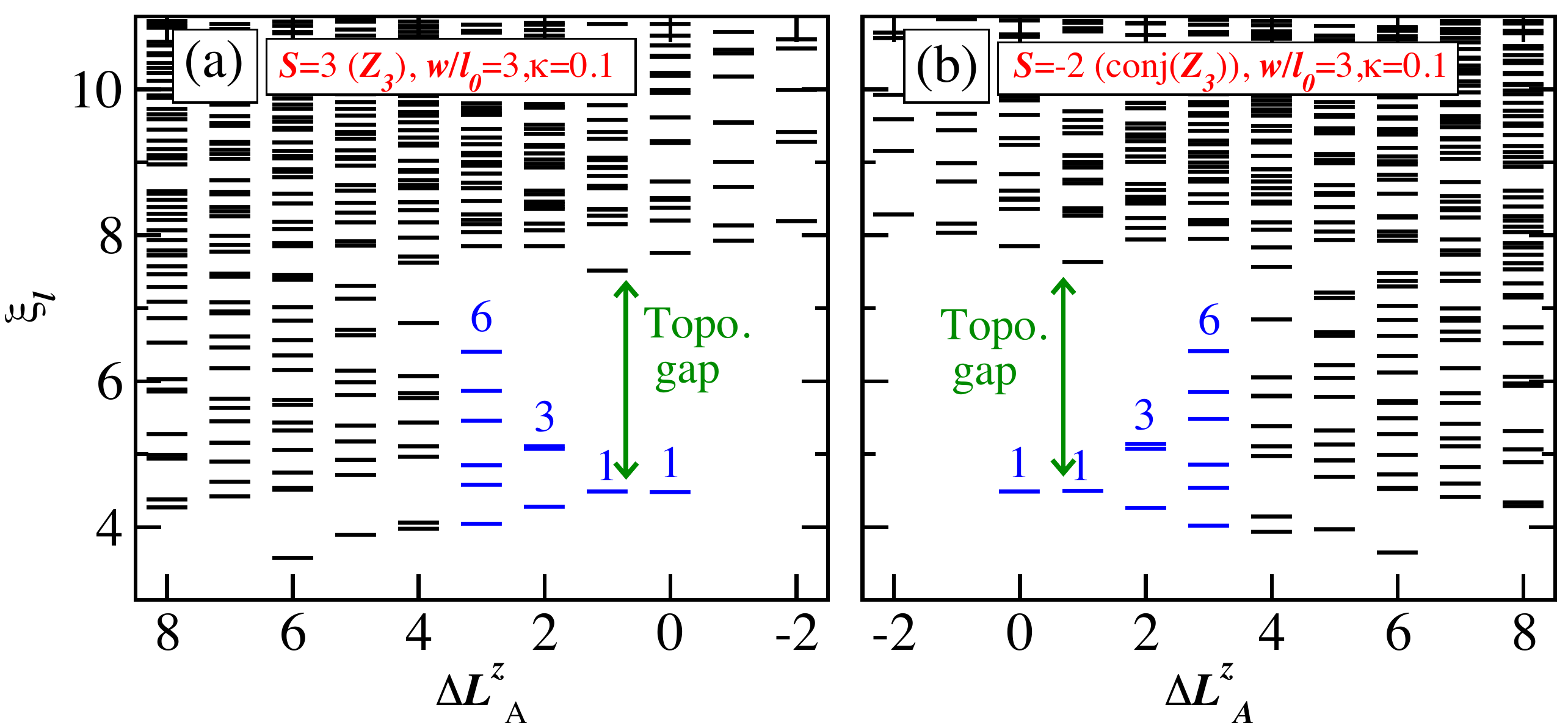}
        \caption{(Color online) Entanglement spectrum for the exact ground state of Eq.~(\ref{Heff}) for  $w/l_0=3$ and $\kappa=0.1$ at (a) $\nu=13/5$ (shift $S=3$) and (b)  $\nu=12/5$ (shift $S=-2$) for $N_\phi=37$. The  counting for the low-lying levels is 1, 1, 3, and 6 up to $\Delta L_A^z=5$, agreeing with  $Z_3$ and conj($Z_3$).   The orbital cuts, using the notation of Ref.~\onlinecite{Li08}, are $P[0|0]$ for $S=3$ and $P[1|1]$ for $S=-2$.  $\Delta L_A^z = L_A^z-(L_A^z)_\mathrm{root}$  where (a) $(L_A^z)_\mathrm{root}=120$ and (b) $(L_A^z)_\mathrm{root}=60.5$. The topological gap is indicated by the green arrow and defined as the difference between the two lowest-lying levels at $\Delta L_A^z=1$ (see Sec.~\ref{sec-gap}).}
   \label{fig:ES}
\end{figure}

\section{Overlap, perturbation theory, and entanglement spectra}
We first investigate whether the system remains in the $Z_3$ RR phase under realistic conditions.  The ground state of Eq.~(\ref{Heff}) is uniform with $L=0$ for the RR shifts for all system sizes up to $N_\phi=37$ for $\kappa\neq 0$ and $N_\phi=42$ for $\kappa=0$ (we have not studied $\kappa\neq 0$ for $N_\phi=42$).  The ground states have $L\neq 0$ for the CF shifts for zero and non-zero $\kappa$, for most system sizes.  
The Bonderson-Slingerland (BS) non-Abelian state for $\nu=12/5$~\cite{Bonderson08} has $L=0$ at $\kappa=0$, but a smaller gap than the RR state~\cite{Bonderson2012}--this behavior remains with $\kappa\neq 0$; see Appendix~\ref{sec:bs}.
Similar qualitative results were recently found in the $\kappa=0$ limit~\cite{ZhuHaldane2015,ZaletelMongArxiv2015}.

Figure~\ref{fig:z3overlaps_H3expValPlanarCoul}(a) presents the overlap between the exact ground state $|\psi\rangle$ of Eq.~(\ref{Heff}) with the model wave functions [$Z_3$  and conj($Z_3$)].  For small $\kappa$, the overlap remains relatively unchanged, but the 12/5 overlap with conj($Z_3$) is larger than the overlap with $Z_3$ at 13/5 for $\kappa \lesssim 0.5$ for all system sizes--the overlap at 13/5 decreases monotonically with $\kappa$ and both overlaps are found to collapse to zero near $\kappa\approx 1$, though some finite-size effects are observed for larger $\kappa$.

Since the overlaps are relatively flat for small $\kappa$, we study the eigenstates obtained in the absence of LL mixing, at $\kappa=0$ (denoted $|\psi_0\rangle$).  We calculate $\langle\psi_0|H^{(3)}(\kappa=0.1)|\psi_0\rangle$, where $H^{(3)}(\kappa)=\sum_{m} V^{(N)}_{\mathrm{3body},m}(w/{\ell_0},\kappa)\sum_{i<j<k}\hat{P}_{ijk}(m)$ [shown in Fig.~\ref{fig:z3overlaps_H3expValPlanarCoul}(b)]--this represents the lowest-order perturbative contribution to particle-hole symmetry breaking induced by LL mixing.   The thermodynamic limit extrapolation of  $\left<\psi_0|H^{(3)}(\kappa=0.1)|\psi_0\right>$ per particle for  $\nu=12/5$ is more than ten times smaller than for 13/5, indicating that LL mixing more severely affects the energetics of 13/5 compared to 12/5.  While the ground-state energies are lowered by the three-body terms, the excited states are lowered as well, reducing the energy gap at 13/5 and increasing the gap at 12/5.  In the inset of Fig.~\ref{fig:z3overlaps_H3expValPlanarCoul}(b), we show that $V^{(1)}_{\mathrm{3body},3}$, $V^{(1)}_{\mathrm{3body},5}$, and $V^{(1)}_{\mathrm{3body},6}$ are the three-body pseudopotentials that  contribute most to particle-hole symmetry breaking between $\nu=12/5$ and 13/5.  The $Z_3$ state has a relative abundance of three-body clustering by construction~\cite{Read99} and  large expectation value of $H^{(3)}(\kappa)$ (not shown), similar to $|\psi_0\rangle$ at $\nu=13/5$.  In contrast, the three-body terms have little effect on 12/5.

Overlaps may depend on short-range physics, so we investigate orbital entanglement spectra~\cite{Levin06,Kitaev06b,Haque07,Zozulya07,Li08,Biddle11}.  If the ground state is in the RR phase, the  counting of the low-lying levels of the entanglement spectra will be related to the SU$(2)_3$ TQFT describing the edge excitations~\cite{Li08}. The  counting of the low-lying levels  for  $\nu=13/5$ and 12/5 for $w/\l_0=3$ and $\kappa=0.1$ (Fig.~\ref{fig:ES}) matches the  counting for $Z_3$ and conj($Z_3$), respectively, (including $\kappa=0$; see Ref.~\onlinecite{ZhuHaldane2015}).

The results above confirm that the ground state of Eq.~(\ref{Heff}) remains in the RR phase under LL mixing.  Further, LL mixing  affects $\nu=13/5$  more than  12/5 and introduces strong particle-hole asymmetry.

\begin{figure}[]
 \includegraphics[width=8.5cm,angle=0]{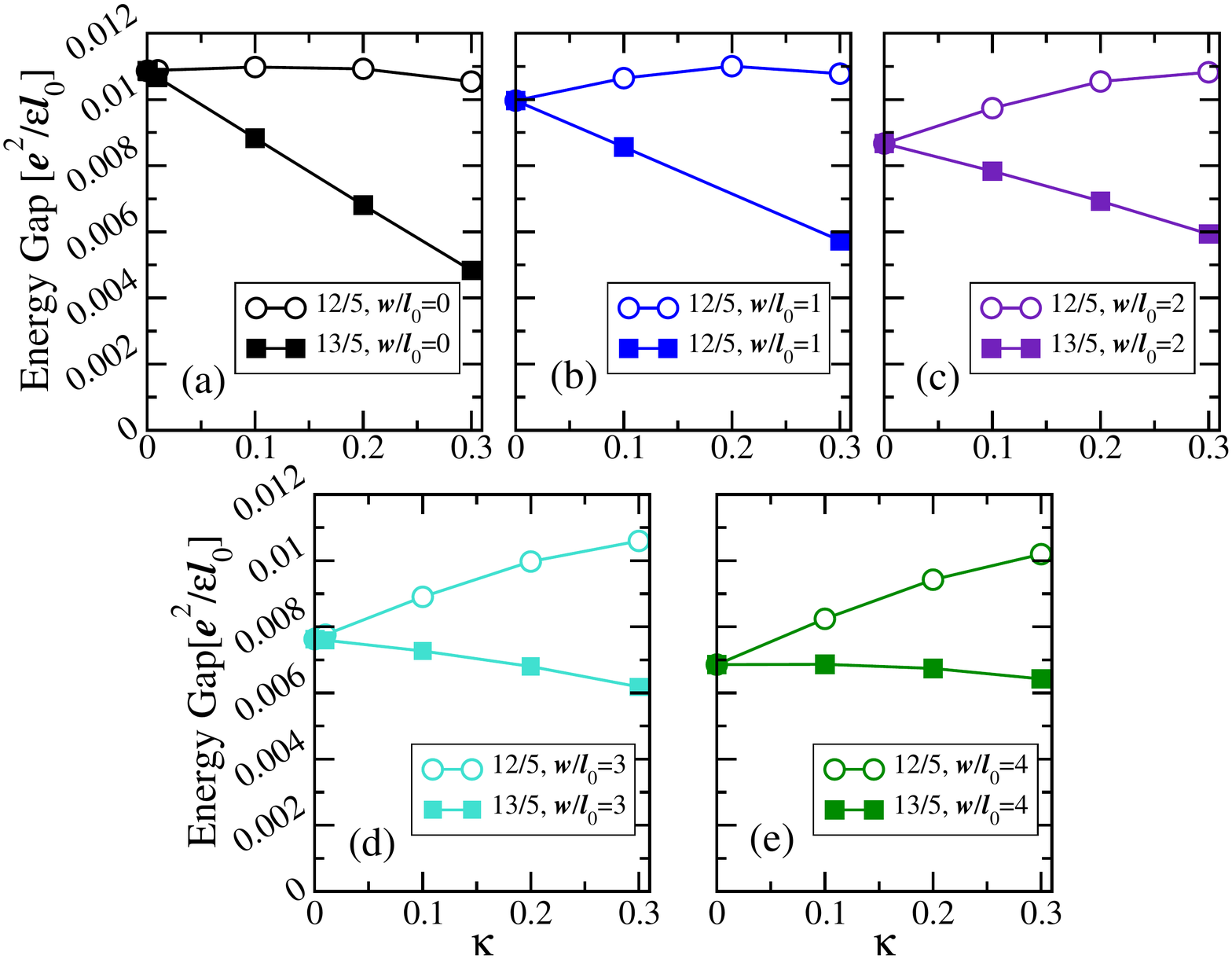}\\
  \includegraphics[width=8.5cm,angle=0]{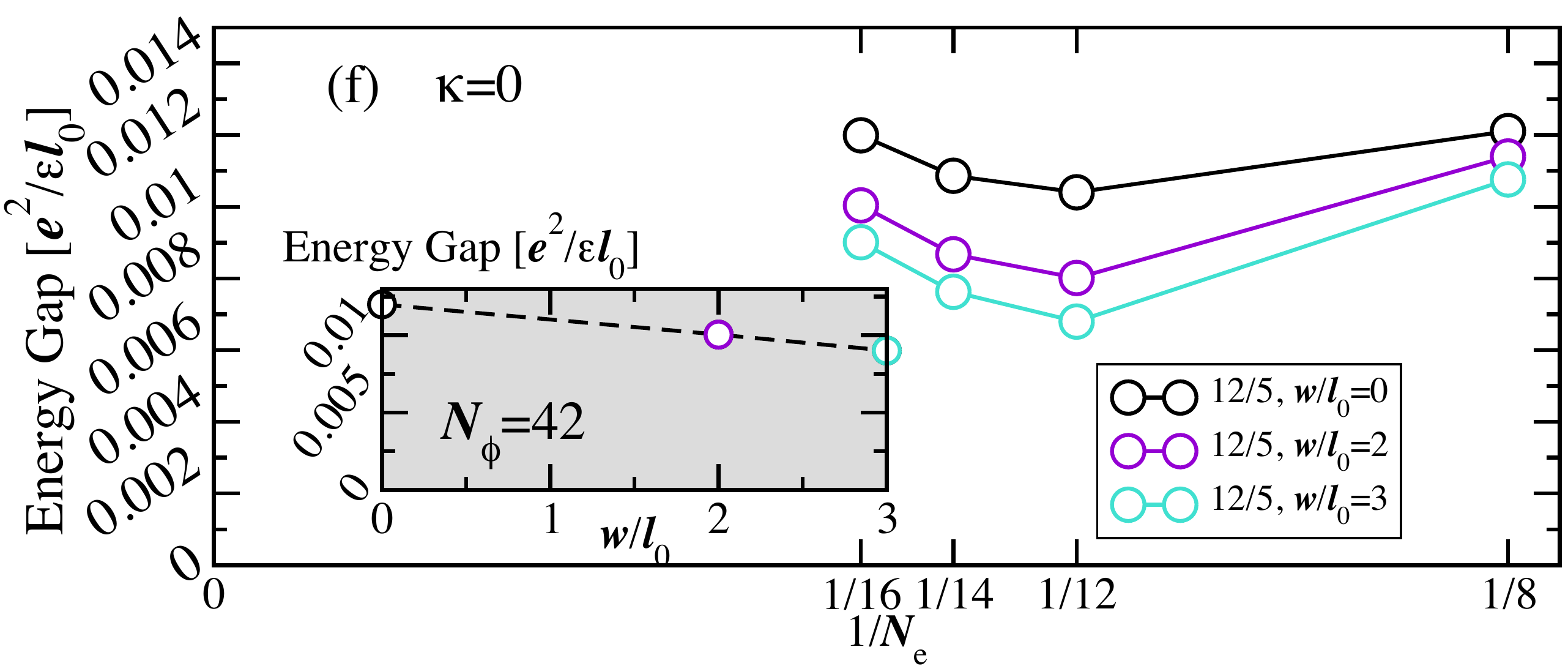}
 \caption{(Color online)  Energy gap for $N_{\phi}$=37 at $\nu=12/5$ and 13/5 for (a)-(e) $w/l_0=0$-$4$.   Similar results are obtained for smaller system sizes.  (f) Width dependence of the  gap for $N_e=$8, 12, 14, and 16 for $\nu=12/5$ for $w/l_0=0$, 2, and 3 and $\kappa=0$.  Inset: The gap as a function of $w/l_0$ at $\kappa=0$ for $N_e=16$  ($N_\phi=42$).  Finite width reduces the gap by approximately 25$\%$ at $w/l_0=3$ relative  to $w/l_0=0$ for the largest system size.  Note the similarities in (f) to Fig. 1(b) in Ref.~\onlinecite{ZhuHaldane2015}.}
   \label{fig:EnergyGapsNphi37w0and3}
\end{figure}

\section{Energy gap and topological gap}
\label{sec-gap}
The neutral gap  is related to the experimentally measured activation gap and  the physical robustness of the FQHE.  It is the difference between the two lowest energies at constant $N_\phi$, if the ground state has $L=0$, otherwise it is taken to be zero. 

Figure~\ref{fig:EnergyGapsNphi37w0and3}(a)-3(e)  show  energy gaps for our largest system  ($N_{\phi}=37$) for $w/l_0=0$-4, respectively.  LL mixing breaks  particle-hole symmetry, producing a larger energy gap for $\nu=12/5$ compared to 13/5. The  gap at $w/l_0\ne0$ for 12/5 increases with $\kappa$, while the 13/5   gap is suppressed (the suppression  is found for all non-aliased system sizes and  values of $w/l_0$; however, an increasing gap at $\nu=12/5$ for non-zero width is only found for the two largest system sizes $N_{\phi}=37$ and 32).  Hence, LL mixing strengthens the 12/5 FQHE for finite $w/l_0$, while weakening 13/5 (strengthening of the FQHE gap with LL mixing does not happen for $\nu=5/2$~\cite{Pakrouski2015}).   

The thermodynamic extrapolation suffers from finite-size effects ($N_{\phi}=12$ and $17$) and aliasing ($N_\phi=27$).  The energy gaps at the remaining $N_{\phi}$ are shown  in Fig.~\ref{fig:EnergyGapsNphi37w0and3}(f).  Without LL mixing,  finite width  decreases the gap from $0.012e^2/\epsilon l_0$ at $w/l_0=0$ to $0.009e^2/\epsilon l_0$ at $w/l_0=3$ [values given are for  $N_{\phi}=42$, shown in the inset of Fig.~\ref{fig:EnergyGapsNphi37w0and3}(f)]. In the limit of small LL mixing, (i.e., high magnetic fields),  it should be possible to observe more robust 12/5 states in narrow quantum wells. 

We expect that the equivalence of various models of finite width demonstrated for $\nu=5/2$~\cite{Pakrouski2015} also holds here. Thus, to determine the effective width $w/\l_0$ corresponding to a certain  experimental device, one would first calculate (for instance, using a Schrodinger-Poisson solver) or measure~\cite{Reichl2015} the square of the absolute value of the electron wave function in the direction perpendicular to the two-dimensional electron gas (2DEG) and determine its variance (as defined in Ref.~\onlinecite{Pakrouski2015}). Then,  $w/\l_0$ should be chosen such that the variance in the ground state of an infinitely deep quantum well of width $w/\l_0$ is the same as in the given experimental sample.

\begin{figure}[t!]
\includegraphics[width=8.5cm,angle=0]{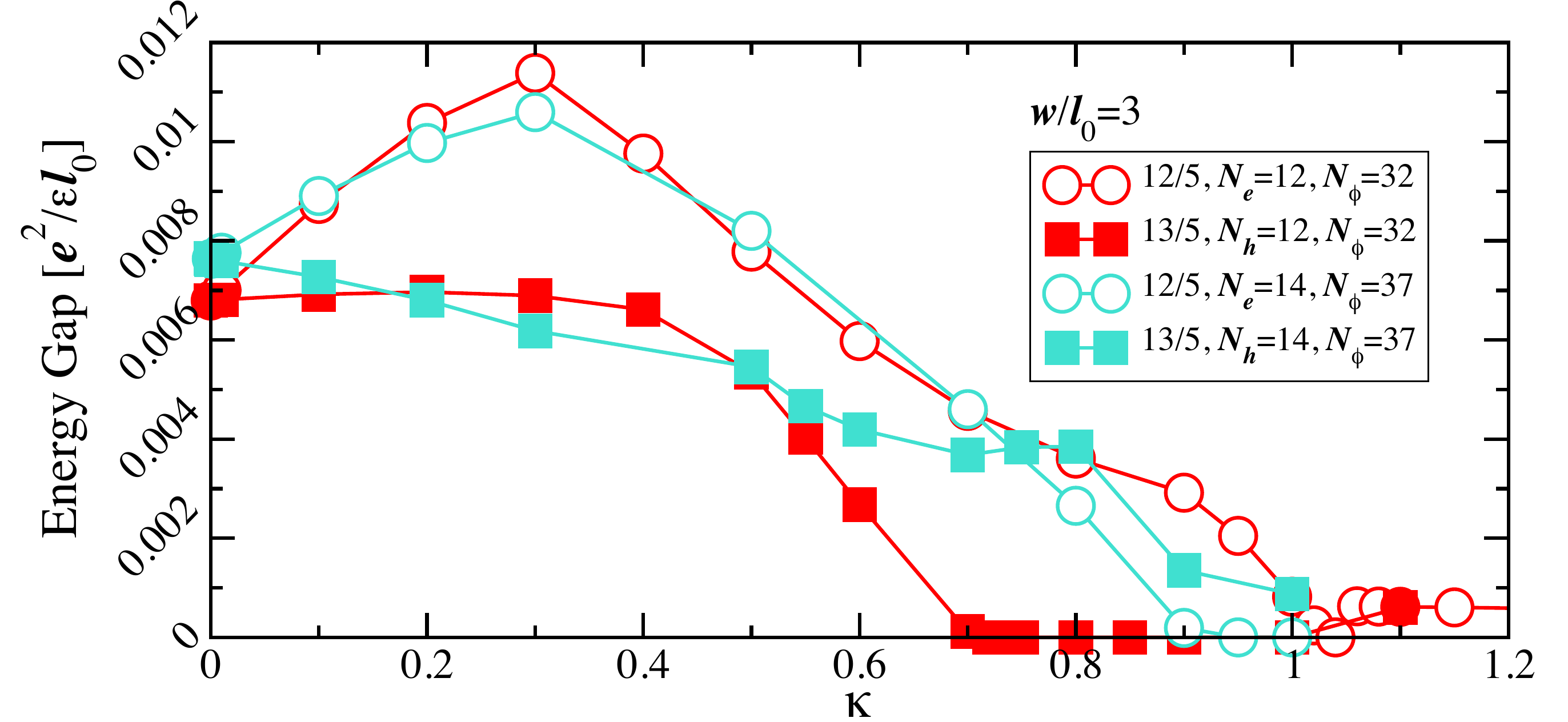}
 \caption{(Color online)  Energy gap for $\nu=12/5$ and 13/5 as a function of $\kappa$ for $w/\l_0=3$ for $N_\phi=32$ and 37.  Energy gap for 12 holes at 13/5 is put to 0 for $\kappa \ge 0.72$, where the ground state has gone through a phase transition into a non-homogeneous state with $L=2$. We note for $\kappa \gtrsim 0.6$ the gap behavior is no longer consistent between system sizes. }
   \label{fig:gapN12N14}
\end{figure}

Figure~\ref{fig:gapN12N14} shows the energy gap as a function of $\kappa$ for $N_\phi=32$ and 37 [12 and 14 electrons (holes) for $\nu=12/5$ (13/5), respectively]  to the experimental value of $\kappa\sim 1.1$ for $w/l_0=3$. All of the sharp features in the $\kappa$-dependence are associated with the change of $L$ in the first-excited states. The behavior of the different system sizes is consistent up to $\kappa=0.6-0.7$ and demonstrates a larger energy gap at 12/5 than at 13/5.  Finite-size effects are observed for larger $\kappa$, which could be a result of our perturbative (in $\kappa$) approach to LL mixing breaking down or the smallness of the energy gap.

\begin{figure}[t!]
        \includegraphics[width=7.5cm,angle=0]{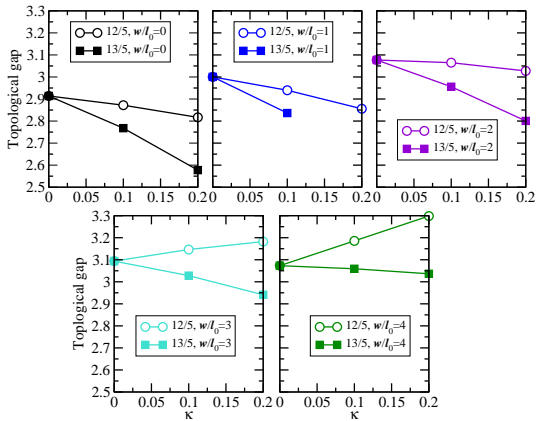}
        \caption{(Color online) Topological gap for 12/5 and 13/5 as a function of $\kappa$ for $w/\l_0=0$-$4$ and $N_{\phi}=37$.}
   \label{fig:topgaps}
\end{figure}

Finally, we investigate the topological gap.  Following Ref.~\onlinecite{Li08}, we define the topological gap as the difference between the two lowest-lying levels in the entanglement spectrum at $\Delta L^z_A=1$  (see Fig.~\ref{fig:ES}). It represents the ``energy difference" between the universal part of the entanglement spectrum, describing the [non-Abelian in the case of RR and conj(RR)] modes and the generic continuum of states.
In Fig.~\ref{fig:topgaps}, we identify two trends: first, the topological gap increases with increased finite width, and second, Landau-level mixing leads to the suppression of the topological gap at 13/5 relative to 12/5 in the same way as observed for the energy gap, giving support to the main conclusion of this work based on a different measure. 

\section{Second versus lowest Landau level}
\begin{figure}[]
        \includegraphics[width=8.5cm,angle=0]{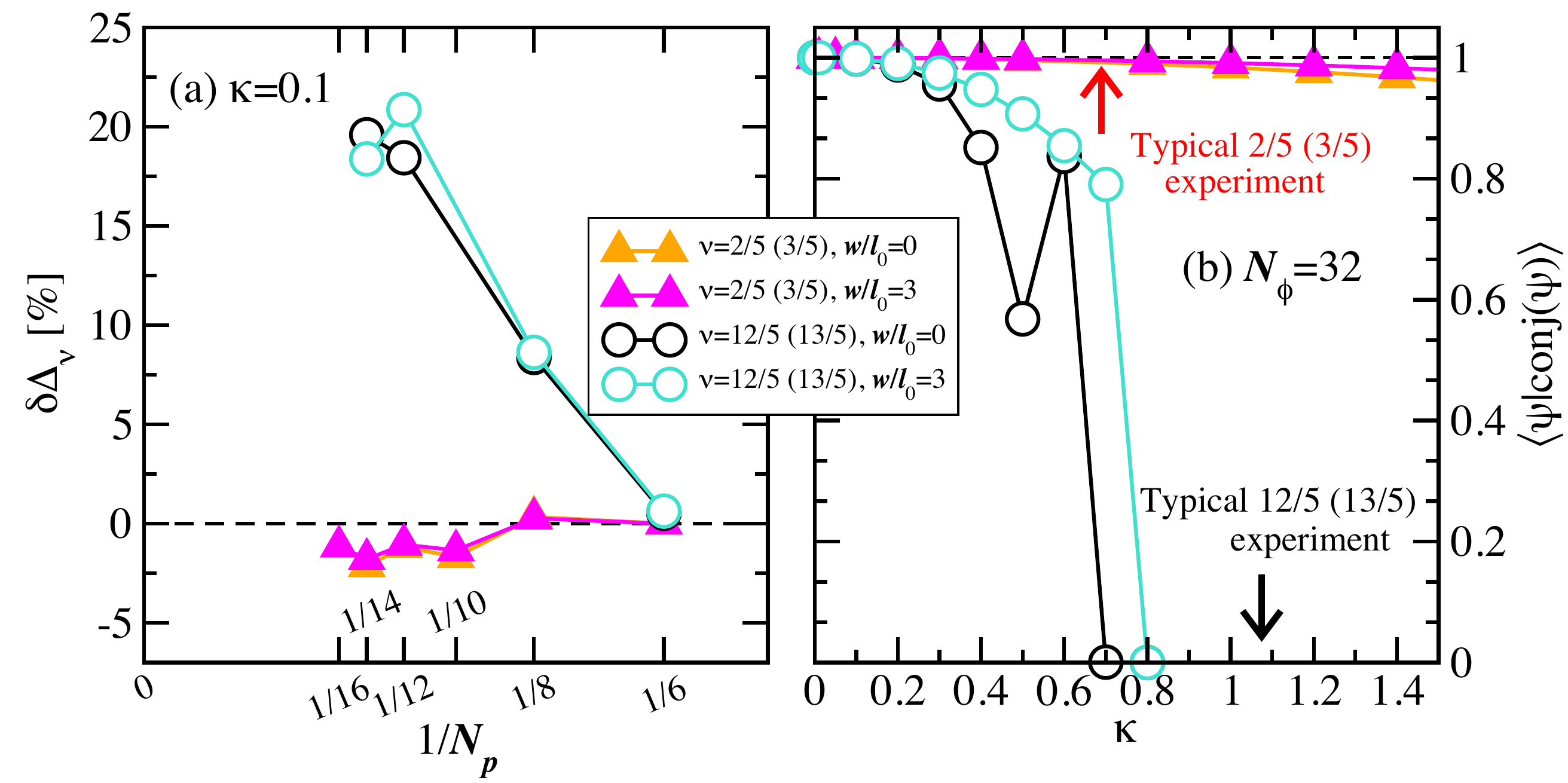}
        \caption{(Color online) (a) Relative  gap difference $\delta\Delta_\nu=(\Delta_\nu-\Delta_{1-\nu})/\Delta_\nu$ (induced by  $\kappa=0.1$) between particle-hole-conjugates at 12/5 (13/5) and 2/5 (3/5). $N_p$ is the number of particles for $\nu=12/5$ and 2/5 or number of holes for $\nu=13/5$ and 3/5.  (b) Particle-hole symmetry breaking [quantified by $\langle \psi|\mathrm{conj}(\psi)\rangle$] in the second LL compared to the lowest LL for $w/l_0=0$ and 3. The system sizes are $N_{\phi}=32$ for $\nu=12/5$ (13/5) and $N_{\phi}=31$ for $\nu=2/5$ (3/5). 
       }
   \label{fig:LLLvsSLL}
\end{figure}

Finally, we compare the second with the lowest LL. In Fig.~\ref{fig:LLLvsSLL}(a), we show the relative energy gap difference induced by LL mixing between $\nu=12/5$ and 13/5 and between $\nu=2/5$ and 3/5 as a function of particle number.  The LL mixing induced difference is much larger in the second LL than in the lowest LL (the sign is also different between the two, with 12/5  strongly favored in the second LL while 3/5 is slightly favored in the lowest LL).  The LL mixing induced gap difference between 12/5 and 13/5 grows with system size and is  likely  a robust feature in the thermodynamic limit.  

We can further quantify the particle-hole symmetry breaking  by calculating the overlap between the exact ground state $|\psi\rangle$ at $\nu=12/5$ (2/5) and the particle-hole conjugate of the exact ground state $|\mathrm{conj}(\psi)\rangle$ at $\nu=13/5$ (3/5).  At $\kappa=0$, this overlap is unity since the two states are particle-hole conjugates.  In Fig.~\ref{fig:LLLvsSLL}(b), particle-hole symmetry is much more strongly broken for the $\nu=12/5$ (13/5) FQHE than for the $\nu=2/5$ (3/5) FQHE. In fact,  particle-hole symmetry is hardly broken at all in the lowest LL [in the lowest LL, $\langle \psi|\mathrm{conj}(\psi)\rangle\gtrsim0.9$ up to $\kappa\sim 2.4$].   This apparent particle-hole symmetry could be a property of the lowest LL  or of the CF-like states in any LL.  

\section{Conclusion}
LL mixing strongly breaks the particle-hole symmetry between  $\nu=12/5$ and 13/5 FQHE in the second LL, but has little effect on  $\nu=2/5$ and 3/5 FQHE in the lowest LL. Our work implies  that the absence of 13/5 FQHE in the second LL is likely a direct consequence of LL mixing effects. This is mainly due to the suppression of the energy gap at $\nu=13/5$ -- the FQHE might simply be too fragile (in terms of energy gap) since LL mixing affects 13/5 more severely than 12/5, and  because in experimental measurements, at constant density, $\kappa$ is  larger at 13/5 compared to 12/5 (since the magnetic field at 13/5 is smaller than at 12/5).  The 12/5 ground state at shift $S=-2$ remains in the non-Abelian parafermionic (conjugate) RR $Z_3$ phase when finite-width and non-zero LL mixing are taken into account extending the validity of previous conclusions~\cite{ZhuHaldane2015,ZaletelMongArxiv2015,Read99,RR2009,Wojs2009} obtained for idealized conditions. We do not  rule out the $\nu=13/5$ FQHE in the $Z_3$ RR phase, but establish that  the 13/5 FQHE is always much weaker than  12/5.  Future experiments with smaller $\kappa$ could show a very weak FQHE at $\nu=13/5$ in extremely high-mobility samples at ultra-low temperatures with a very small activation energy.

\acknowledgements{M.T. and K.P. were supported by the Schweizerischer Nationalfonds zur F{\"o}rderung der Wissenschaftlichen Forschung, National Centre of Compentence in Research -- Quantum Science and Technology, the European Research Council, ERC Advanced Grant SIMCOFE and by Microsoft Research.  M.R.P. was supported by the National Science Foundation under Grant No. DMR-1508290, the Office of Research and Sponsored Programs at California State University Long Beach, and the W. M. Keck Foundation. Y.L.W. and S.D.S. were supported by Microsoft Research and LPS-MPO-CMTC.  This work was supported by a grant from the Swiss National Supercomputing Centre (CSCS) under Project IDs s395 and s551. The authors are grateful to C. Nayak, R. Mong, R. Morf, M. Zaletel, and also P. Bonderson and S. Simon for many helpful discussions.}

\appendix

\section{Energetics at the Bonderson-Slingerland shift}
\label{sec:bs}

\begin{figure}[t!]	
  \includegraphics[width=7.5cm,angle=0]{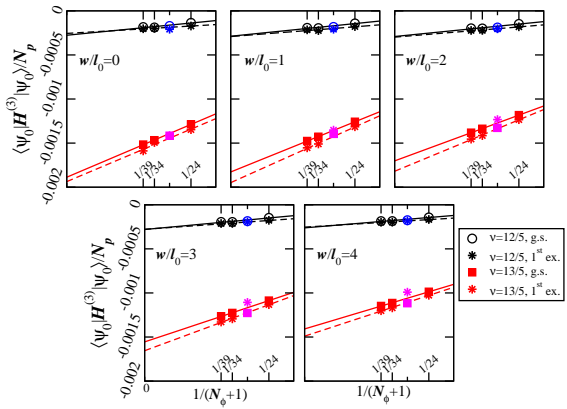}
    \caption{(Color online) 
    Expectation values of the three-body terms per particle $N_p$ of Eq.~(\ref{Heff})    for $\kappa=0.1$ and $w/l_0=0,1,2,3$, and 4 evaluated for the ideal Coulomb ground and first-excited states at 12/5  and 13/5, respectively, as a function of inverse Landau level degeneracy [$1/(N_\phi+1)$] extrapolated to the thermodynamic limit. The only difference from Fig.~\ref{fig:z3overlaps_H3expValPlanarCoul}(b) is that the ground state at the Bonderson-Slingerland shift and its particle-hole conjugate ($S=2$ and $S=1/3$) were taken instead of the ones for the Read-Rezayi and conjugate shifts.  Lines  represent linear extrapolations, excluding the $N_{\phi}=28$ data point which appears to behave differently from all other system sizes [these points are indicated in blue (12/5) and magenta (13/5), respectively.]}  
   \label{fig:suppBS}
\end{figure}

In this appendix, we explore the perturbative change in the FQHE gap Landau-level mixing induced at the shifts corresponding to the Bonderson-Slingerland (BS) state  and its corresponding particle-hole conjugate.  Shown in Fig.~\ref{fig:suppBS} are the expectation values of the three-body terms of our effective Hamiltonian [Eq.~(\ref{Heff})] for the ground and first-excited states [the results are presented in the same way as in Fig. 1(b)].  Both the ground and excited states reduce their energy by approximately the same amount at 12/5.  For 13/5, the energy of the excited state is reduced significantly more than that of the ground state meaning that the gap of 13/5 is reduced, whereas the gap of 12/5 remains relatively constant.

\section{Robustness of the composite fermion states for the 2/5 and 3/5 FQHE under Landau level mixing}

\begin{figure}[t!]
        \includegraphics[width=7.5cm,angle=0]{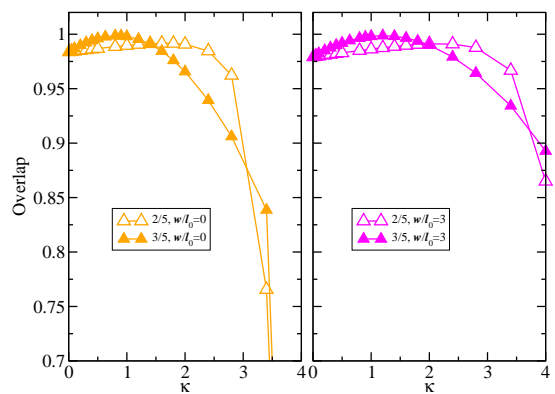}
        \caption{(Color online) Overlap between the realistic ground state and the ground state of the hardcore ($V_1\neq 0$ and $V_m=0$ for all other $m$) Hamiltonian for $N_{\phi}=31$. w=0 (left panel) and w=3 (right panel).
        }
   \label{fig:2fand3fOverlaps}
\end{figure}

To further characterize the evolution of the states in the lowest Landau level, we approximate the CF-like states at 2/5 and 3/5 with the exact ground state of a ``hard-core" model Hamiltonian with $V_1\neq 0$ and all other $V_m=0$ at $N_\phi=5N_e/2-4$ and $N_\phi=5N_e/3+1$, respectively.  This Hamiltonian produces the $1/m$ Laughlin state exactly as the zero-energy ground state for $N_\phi = m(N_e-1)$ and produces ground states with large overlaps ($>0.99$) with CF wave functions for filling factor $\nu=n/(2pn+1)$ at the appropriate flux as checked via Monte Carlo.  As shown in Fig.~\ref{fig:2fand3fOverlaps}, the overlap remains stable under Landau-level mixing and only starts to significantly decrease around $\kappa=3-4$, well beyond the typical experimental values.
 
It is an open question whether the observed robustness of the FQH states at 2/5 and 3/5 is due to their CF-like nature or to the specific form of the effective interaction in the lowest Landau level.


%

\end{document}